\documentclass{egpubl}
\usepackage{eg2017}

 \ShortPresentation      
 \electronicVersion 

\ifpdf \usepackage[pdftex]{graphicx} \pdfcompresslevel=9
\else \usepackage[dvips]{graphicx} \fi

\PrintedOrElectronic

\usepackage{t1enc,dfadobe}

\usepackage{egweblnk}
\usepackage{cite}

\title["Synchronize" to VR Body: Full Body Illusion in VR Space]
{"Synchronize" to VR Body: Full Body Illusion in VR Space}

\author[P. Xiong, C. Sun \& D. Cai]
{\parbox{\textwidth}{\centering Peikun Xiong, Chen Sun and Dongsheng Cai
	}
	\\
	{\parbox{\textwidth}{\centering University of Tsukuba, Graduate School of Systems and Information Engineering, Japan
		} 
	}
}

\usepackage{siunitx}
\begin{document}
	
	 \teaser{
	  \includegraphics[width=.9\linewidth]{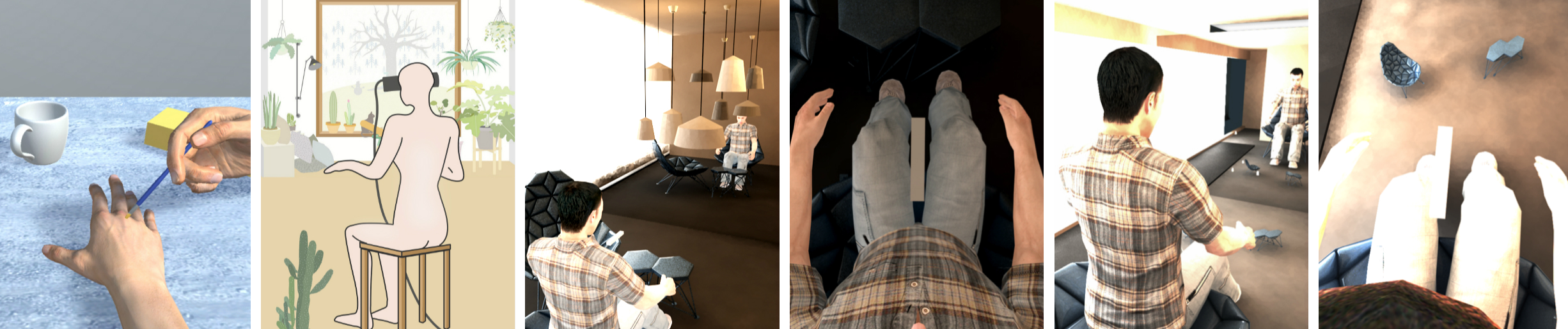}
	  \centering
	   \caption{From left to right, the visual stimulus input we used in VR RHI (Rubber hand illusion) tests, the appearance of a participant in FBI (Full Body Illusion) tests, and full body illusions applied to different size bodies varying from 140-500cm.}
	 \label{fig:1}
	 }

	\maketitle	
	
	\begin{abstract}
		Virtual Reality (VR) becomes accessible to mimic a "real-like" world now. People who have a VR experience usually can be impressed by the immersive feeling, they might consider themselves are actually existed in the VR space. Self-consciousness is important for people to identify their own characters in VR space, and illusory ownership can help people to "build" their "bodies". The rubber hand illusion can convince us a fake hand made by rubber is a part of our bodies under certain circumstances. Researches about autoscopic phenomena extend this illusory to the so-called full body illusion. We conducted 3 type of experiments to study the illusory ownership in VR space as it shows in Figure ~\ref{fig:1}, and we learned: Human body must receive the synchronized visual signal and somatosensory stimulus at the same time; The visual signal must be the first person perceptive; the subject and the virtual body needs to be the same height as much as possible. All these illusory ownerships accompanied by the body temperature decreases, where the body is stimulated.
		
		
	\end{abstract}
	

	\section{Introduction}
	
	William James \cite{Jam90} developed a theory of two selves. The first one is the "Me" self, and the second one is the "I" self. The "me" self refers to the aspects of someone that come from that person's experiences. The "me" self can break down into three sections, The Material Self, The Social Self, and The Spiritual Self. The "I" self is categorized into the thinking self. The Pure Ego was the name given to the "I" self. James also thinks our body is the innermost part of the material self. Gallagher's \cite{gal00} studies between philosophy of mind and the other cognitive science that are focused on two aspects of self. They are the minimal self and the narrative self. The minimal self is the procedure of human brain processes and ecologically embedded body. He considered sense of ownership is an integration of information (i. e. visual, tactile, somatosensory etc.) processing in human brains. According to these various studies, we know the illusory ownership can be evoked in some conditions that the other objects can be recognized as the subjects' own body parts by themselves. Recently, due to the fast development of both computer graphic technologies and computer hardware performances, VR has gained more attention now. It can replicate an environment to allow the users achieving an immersive experience. We have studied approaches and methods that can evoke the illusory ownership in the VR environment. Especially, we use the Head Mount Display (HMD) to evoke illusory ownerships with various conditions.

	\section{Previous researches}
	
	Botvinick and Cohen reported a phenomenon that called Rubber Hand Illusion (RHI) \cite{BC98}, which a rubber hand placed in front of a healthy subject can misunderstand it as his or her own hand under certain circumstances. Many researches on RHI have been conducted recently. Some researches replaced the rubber hand with other objects such as: a table \cite{AR03}, a rake, a curtain, and another stick-shape objects \cite{HP10}. They all made it clear that the subjects felt if these objects were the part of their own bodies, but the other studies (using stick-shape object \cite{TH05}) also pointed out that the non-hand-shape objects cannot evoke the RHI. 
	
	Full Body Illusion (FBI) is based on the research of human brain disease that called Autoscopic Phenomena. The patients have a strong illusion of the ownership of other human bodies. However, the FBI can also occur in healthy humans. It possesses a similar arousal method to the RHI. The research by Petkova \& Ehrsson \cite{PE08} and Slater et al. \cite{SPES09} used both a first-person visual perceptive of the mannequin and tactile stimulus to the test subjects. The report evaluates the subjects' illusion using verbal questionnaires. Lenggenhager et al. \cite{LMB09} and Aspell et al. \cite{ALB09} tries some tests using a third person visual perceptive input, but the subjects still felt the illusion by using the same evaluation methods. 
	
	In most of previous researches, the illusory ownership was evaluated by the investigations about how strong the subjects can feel the illusory ownership. However, we think oral investigations are still unconvincing. With the other methods, electromyography (EMG) was used to measure the difference of muscle activities when a strong blow to the fake hand \cite{THA*12}. They also verified that the EMG difference and strength of illusory ownership are positively related. Furthermore, in \cite{AR03}, the human skin resistivity was also measured if an illusory ownership is evoked. He/she is easy to sweat when he/she is nervous or feels fear. So they made an assumption that skin resistance can vary when illusory ownership was evoked.

	\section{RHIs and FBIs measured by body temperature}
	
	In this paper, we evaluated the strength of illusory ownership by measuring the body temperature, which based on the research of Moseley et al. \cite{MOV*08}. When the RHI was evoked, the subjects' hand temperatures were dropped.
	
	\subsection{RHI}
	
	First, we tried to evoke the RHI by using a virtual hand displayed in a HMD. There were 2 group of participants. Here, 8 participants (2 females and 6 males, $mean \pm $SD age = $24.5 \pm $1.59 years) were tested in RHI trial (Trial 1) and RHI control trial (Trial 2) in actual world base on the experiment of Moseley et al. \cite{MOV*08}. The other 8 participants (3 females and 5 males, $mean \pm $SD age=$25 \pm $2.27 years) were tested in RHI in VR space trial (Trial 3). None of them have any cognition diseases and the knowledge about RHI. We were keeping the indoor temperature under $25^\text{o}$C. The rubber hand and virtual hand displayed in HMD are the right hand.
	
	\textbf{Trial 1:} Participants sat with their right hand resting on a table, and we set a block to prevent the participants seeing their own hand. A rubber right hand was place in front of the participants and a towel was placed over the participants' shoulder, hand, and the end of the rubber hand as if the rubber hand was connected to the participants' shoulder. Then we use two paintbrushes to stroke the rubber hand and the participants' subject hand synchronously. 
	
	\textbf{Trial 2:} The rubber hand was removed. The participants were instructed to look at the position where the rubber hand was placed in Trial 1. The other conditions remain unchanged.
	
	\textbf{Trial 3:} Participants sat and wore a HMD, where a virtual hand resting on a virtual table are displayed as they are in front of the participant. A paintbrush stroking animation is added on the virtual hand with the same stroking pattern to the previous trials. The other conditions remain unchanged. Paintbrush stroking in actual world keep synchronized to the animation in VR space.
	
	\textbf{Method:} For each trial, the test lasts 5 minutes and the hand's temperature was taken from a single spot on each participant's hand by a non-contact thermometer in every 30 seconds, which gave a total of 11 temperature readings per participant.
	
	\begin{figure}[htb]
		\centering
		\includegraphics[width=.8\linewidth]{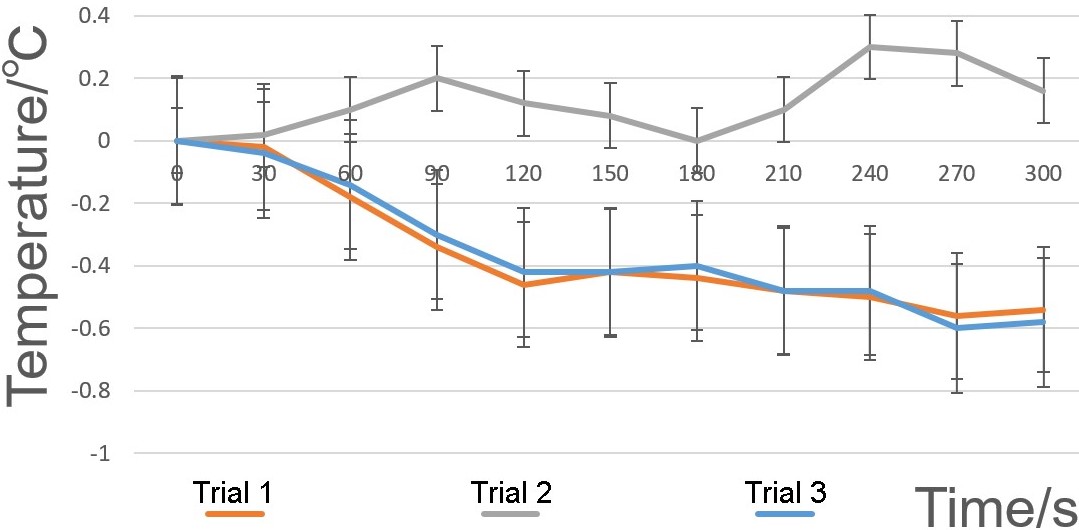}
		\caption{\label{fig:2}
			The average hand temperatures of all subjects in Trial 1, 2, and 3.}
	\end{figure}
	
	As we expected, in Trial 1 and Trial 3, all participants' hand temperatures decreased about 0.6 degrees as illustrated in Figure ~\ref{fig:2}. The T-test of the average temperature between Trial 2 (control) and Trial 3 (VR RHI) is [t (10, 0.005) = 3.16, P = 0.0002 < 1\%]. It has a significant difference. However, the T-test of Trial 1 (RHI) and Trial 3 (VR RHI) is [t (10, 0.025) = 2.228, P = 0.4592 > 5\%]. There is no significant difference between Trail 1 and Trial 3. Thus, we can safely conclude that RHI in VR space (Trial 3) and the RHI in actual world are related. The RHI can be also evoked in VR space and the skin temperature decreases can be observed.
	
	\subsection{FBI}
	
	According to \cite{LMB09} and \cite{PE08}, they evaluated the strength of illusory ownership in both First Person Perceptive (1pp) and Third Person Perceptive (3pp). In their tests, the subjects were asked if they felt the body they are observing were their own body or the mannequin stand in front of them were themselves. We considered, if there was a contradiction between 1pp and 3pp which can affect the physical body temperature decreases and it may not affect the result of investigation.
	
	Here, 5 participants (1 female and 4 males, $mean \pm $SD age = $24.5 \pm $1.59 years) were participant in both 1pp and 3pp experiment. None of them have any cognition diseases and the knowledge about RHI and FBI. We keep the indoor temperature under $25^\text{o}$C. For each participant, we provided both a sleeveless shirt with two symmetrical holes on its back (the same level as shoulder blade) and a shorts to the knee with two symmetrical holes, in order to measure temperatures in 6 spots (four on the limbs and two on the back). Then, the participants sat and wore the HMD. 
	
	\textbf{1pp experiment:} Participants sat and could see a seated virtual body as if the participants see their own body through the HMD screen. The participants' chest was stroked while the virtual body's chest was stroked either synchronously or asynchronously. 
	
	\textbf{3pp experiment:} Participants sat and could see a virtual mannequin as if the mannequin was sitting in front of them through the HMD. The participant's chest and the mannequin's chest was stroked either synchronously or asynchronously. 
	
	\textbf{Method:} Two experiments were performed in order. Each experiment has two trials, a synchronous and an asynchronous trial. For each trial, stokes lasted 5 minutes and the body temperatures were taken from the 6 spots in order by a non-contact thermometer in every 30 seconds, which gave 11 body temperatures per spot, 66 body temperature measurements at one trial per participant. 
	
	\begin{figure}[htb]
		\centering
		\includegraphics[width=.9\linewidth]{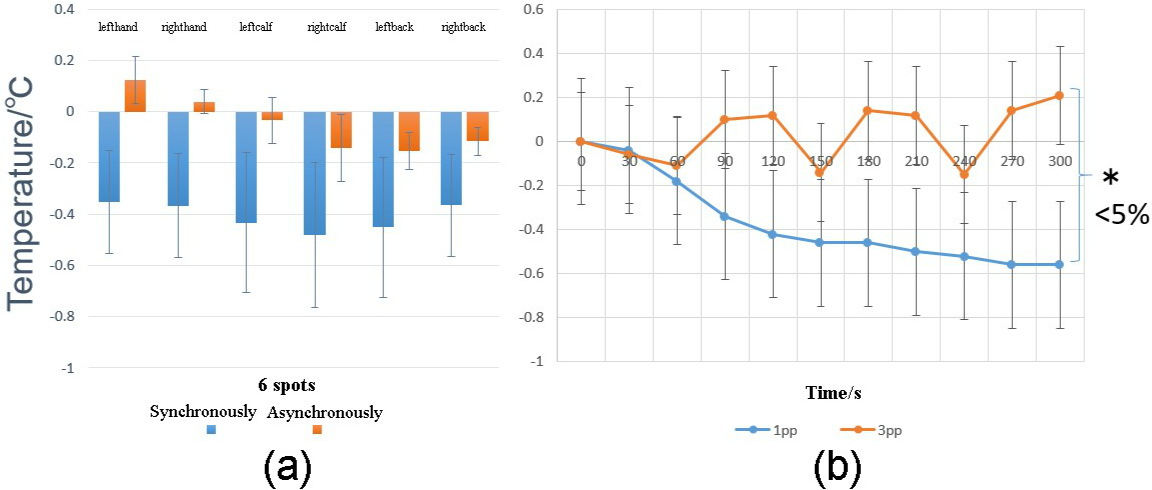}
		\caption{\label{fig:3}(a) The temperature comparison between synchronous and asynchronous stroke in 6 spot. (b) The average hand temperature changing between 1pp and 3pp synchronous trial.}
	\end{figure}
	
	In 1pp experiment, the body temperatures variations are showing in Figure ~\ref{fig:3}. (a). We set the synchrony and the temperatures in different spots to be two factors. The average body temperatures at each symmetrical spots on the left and right were evaluated by ANOVA test. The significant temperature differences in the symmetrical left and right spots such as: the left hand and the right hand [F (2,4) = 0.001, P = 0.982 > 5\%], the left calf and the right calf [F (2,4) = 0.001, P = 0.687 > 5\%], and the left back and the right back [F (2,4) = 0.001, P = 0.778 > 5\%] are all rejected. On the other hand, when we assume the synchronicities are the significant factors, the all significant average temperature differences on the hand [F (2,4) = 0.019 < 5\%], the calf [F (2,4) = 0.005 < 5\%], and the back [F (2,4) = 0.015 < 5\%] cannot be rejected. Therefore, we can safely conclude that a similar mechanism of human brain functions can work on both RHI and FBI. Furthermore, when the illusory ownership is evoked, a consistent temperature dropping can be observed. 
	
	Next, we illustrated the average hand temperature variations of both 1pp and 3pp synchronous trials by using T-test as indicated in Figure ~\ref{fig:3}. (b). There is a significant difference [t (10, 0.005) = 3.169, P = 0.0008<1\%]. We only utilized the temperature on the hand, because we have already found that the significant temperature differences over the 6 spots are rejected. For 1pp experiment, there was a significant difference in hand temperature decreases [F (2,4) = 0.001, P = 0.019 < 5\%]. But for 3pp experiment [F (2,4) = 1.861, P = 0.198 > 5\%], the significant temperature decreases was rejected. But, the previous researches by \cite{LMB09} and \cite{ALB09} claims that they did evoked the FBI in 3pp and they did not show any physical evidence whether the FBI is evoked or not. We observed these physical evidences only in 1pp experiment. Therefore, we considered the actual systemic illusion occurred on both physical body and mental spirit only in the first person perceptive.
	
	\section{FBI for different size virtual bodies}
	
	As some other researches suggest FBI can be affected by the authenticity of virtual mannequin. Banakou et al. \cite{BGS13} tested with bigger and smaller virtual mannequins. In their tests, the subjects gave positive oral answers on the illusory ownership for the both conditions. Here, we intended to accept more explicit responses from the subjects varying the body sizes. 
	
	There were 10 participants (all male, $mean \pm $SD age = $25.5 \pm $1.25 years) with the average height 175 cm. None of them have any cognition diseases and the knowledge about RHI and FBI. We kept the indoor temperature under $25^\text{o}$C. And prepared 4 different size virtual bodies with the height of 140cm, 175cm, 250cm and 500cm. Using each different height body, the participants wore the HMD, and were provided 5 minutes to adapt to their new bodies.
	
	\textbf{140cm experiment:} A 140cm height mannequin was displayed sitting in front of a mirror as participants could see themselves in the mirror. There are two trials, each participant's chest was stroked while the mannequin's chest was stroked either synchronously or asynchronously. In the end of each trial, participants were asked if they felt the ownership to the mannequin.
	
	\textbf{175cm experiment:} A 175cm height mannequin was displayed. The other procedures were identical to 140cm experiment.
	
	\textbf{250cm experiment:} A 250cm height mannequin was displayed. The other procedures were identical to 140cm experiment.
	
	\textbf{500cm experiment:} A 500cm height mannequin was displayed. The other procedures were identical to 140cm experiment.
	
	\textbf{Method:} The method of these four experiments are the same as the previous FBI tests except that we only measure the hand temperatures.

\begin{figure}[htb]
		\centering
		\includegraphics[width=0.9\linewidth]{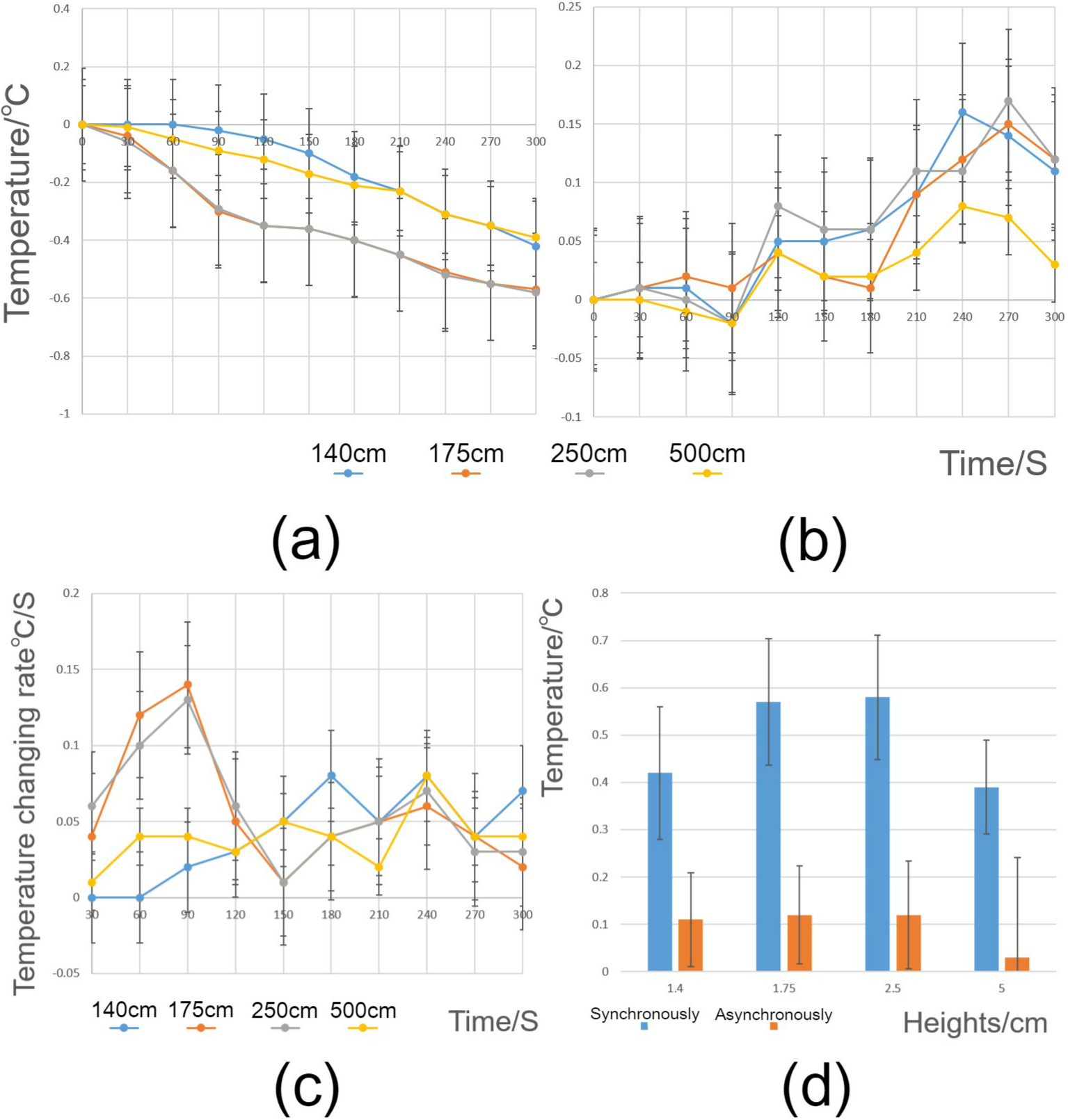}
		\caption{\label{fig:4}The average temperature decreases between the (a) synchronous and (b) asynchronous trials. (c) The temperature changing rate of 4 heights under synchronous trials. (d) the amplitude of body temperature decreases between synchronous and asynchronous trials.}
	\end{figure}

	As it shows in Figure ~\ref{fig:4}. (a) and (b), we observed the average temperature decreases only in the synchronous trials. There is a significant temperature difference between Figure ~\ref{fig:4}. (a) and (b) [F (1,36) = 38.399, P = 0.0001 < 0.1\%]. As the results of four different height mannequins in Figure ~\ref{fig:4}. (a), all the temperature measurements of four heights in the synchronous trials were decreased, and we set the heights as the main effect can be rejected [F (3,36) = 0.08, P = 0.971 > 5\%]. The consequence of these four experiments we conducted is consistent with the previous research \cite{BGS13}.
	
	However, although the factor of mannequin heights had no significant effect on the body temperature, we found that the decrease rate of the body temperature is significantly different varying the different height mannequins. In Figure ~\ref{fig:4}. (c), on the synchronous trials, the 175cm and the 250cm trials have a large variation from time 30s to 120s. Thus, we classified the 175cm and the 250cm trials to be group 1 and the other two heights to be group 2. Then we found there is a significant difference between two groups [F (1,18) = 4.89, P = 0.042 < 5\%]. Furthermore, as it shows in Figure ~\ref{fig:4}. (d), on the asynchronous trials, the temperature did not change significantly and it can be tolerated as normal skin temperature changes. On the synchronous trials, the body temperature decreases for each height are, respectively, $0.42^\text{o}$C, $0.57^\text{o}$C, $0.58^\text{o}$C and $0.39^\text{o}$C. Also, the temperature decrease amplitude of group 1 is larger than the group 2 can be observed in Figure ~\ref{fig:4}. (d). And the difference in body temperature decreases in these two groups with a significant difference level [F (1,38) = 8.10, P = 0.035 < 5\%].As the result shows the temperature changing rate and the temperature decreases amplitude between these two group are different. Thus, we can say the group 1 has a stronger feeling of illusory ownership. Also the 175cm height is the average height of the participants, and one of strongest feeling of illusory ownership was evoked in this height. Therefore, the authenticity of virtual mannequin heights can affect the strength of illusory ownership.
	
	\section{Conclusions}
	
	Illusory ownership is evoked in VR space by using HMD. The evoked FBI can be measured by using the body temperature decreases. The FBI can only be evoked in 1pp tests. We consider the evocation conditions of RHI and FBI are identical, and they can be summarized into three conditions as follows:
	
	1. Human body must receive both the synchronized visual and somatosensory signals in temporal coincidence;
	
	2. The visual signal must be the first person perceptive (1pp);
	
	3. The subject and the virtual body needs to be the same height as much as possible.
	
	The FBI can or may strongly induce or enhance the strong sense of unity with the virtual characters or avatars in VR spaces. The FBI can be used for teleoperation of avatar robots and can enhance the immersive feeling in VR spaces such as those in 4D cinemas. Manipulation of robots with different sizes is one of the unresolved problems, and FBI may help to resolve this problem. However, further researches are necessary to answer these questions. Also we will conduct some case studies to help the VR to be more comfortable for the sensitive users and collaborate with other fields (neuroscience e.g.) by using openBCI device to find out the impact on long-term VR users in the future work. 
	

	

	\bibliographystyle{eg-alpha-doi}
	
	\bibliography{egbibsample}
	

\end{document}